\RequirePackage{fix-cm}
\documentclass[smallextended]{svjour3}       
\smartqed  
\usepackage{graphicx}
\usepackage{amssymb,amsmath,mathptm,mathptmx}
\usepackage{bm,enumerate}
\usepackage{textcomp}
\usepackage[colorlinks=true,citecolor=blue,urlcolor=blue]{hyperref}
\usepackage[mathletters]{ucs}
\usepackage[utf8x]{inputenc}
\usepackage{cite}
\newcommand{\be}{\begin{equation}} \newcommand{\ee}{\end{equation}}
\newcommand{\ba}{\begin{aligned}} \newcommand{\ea}{\end{aligned}}

\newcommand{\Ab}[1]{ \left| #1 \, \right|} 

\DeclareMathOperator{\Tr}{tr}
\DeclareMathOperator{\Hull}{Hull}

\newcommand\bigforall{\mbox{\Large $\mathsurround=0pt\forall$}}
\newcommand\bigexists{\mbox{\Large $\mathsurround=0pt\exists$}}

\newtheorem{trm}{Theorem}

\begin{document}

\title{Quantum randomness protected against detection loophole attacks}


\author{Piotr~Mironowicz$^{1,2}$    \and Gustavo~Ca\~nas$^{3}$ \and Jaime~Cari\~ne$^{4,5}$ \and Esteban~S.~G\'omez$^{4}$ \and Johanna~F.~Barra$^{4,5}$ \and Ad\'an~Cabello$^{6,7}$ \and Guilherme~B.~Xavier$^{8}$ \and Gustavo~Lima$^{4,5}$ \and Marcin~Paw{\l}owski$^{2,9}$
}


\institute{Piotr~Mironowicz\\
\email{piotr.mironowicz@gmail.com}\\
\\
$^1$ Department of Algorithms and System Modeling, Faculty of Electronics, Telecommunications and Informatics, Gda\'{n}sk University of Technology, Gda\'{n}sk 80-233, Poland.\\
$^2$ Institute of Theoretical Physics and Astrophysics, National Quantum Information Centre, Faculty of Mathematics, Physics and Informatics, University of Gda\'nsk, Wita Stwosza 57, 80-308 Gda\'nsk, Poland.\\
$^3$ Departamento de F\'{\i}sica, Universidad del Bio-Bio, Av. Collao 1202, Concepci\'on, Chile.\\
$^4$ Departamento de F\'{\i}sica, Universidad de Concepci\'on, 160-C Concepci\'on, Chile.\\
$^5$ Millennium Institute for Research in Optics, Universidad de Concepci\'on, 160-C Concepci\'on, Chile.\\
$^6$ Departamento de F\'{\i}sica Aplicada II, Universidad de Sevilla, E-41012 Sevilla, Spain.\\
$^7$ Instituto Carlos~I de F\'{\i}sica Te\'orica y Computacional, Universidad de Sevilla, E-41012 Sevilla, Spain.\\
$^8$ Institutionen f\"or Systemteknik, Link\"opings Universitet, 581 83 Link\"oping, Sweden.\\
$^9$ Instytut Fizyki Teoretycznej i Astrofizyki, Uniwersytet Gda\'{n}ski, PL-80-952 Gda\'{n}sk, Poland.
}

\date{Received: date / Accepted: date}

\maketitle

\begin{abstract}
Device and semi-device independent quantum randomness generators (DI- and SDI-QRNGs) are crucial for applications requiring private randomness. However, they are vulnerable to detection inefficiency attacks and this limits severely their usage for practical purposes. Here, we present a method for protecting SDI-QRNGs in prepare-and-measure scenarios against detection inefficiency attacks. The key idea is the introduction of a blocking device that adds failures in the communication between the preparation and measurement devices. We prove that, for any detection efficiency, there is a blocking rate that provides protection against these attacks. We experimentally demonstrate the generation of private randomness using weak coherent states and standard avalanche photo-detectors.
\end{abstract}


\section{Introduction}
\label{intro}


Private random numbers are essential for multiple applications, including, but not limited to, cryptography and digital rights management. Private random numbers are those the user is sure no one else had access to. However, random numbers produced from classical processes may be predictable and therefore not private. One solution is using quantum random number generators (QRNGs) based on the intrinsic uncertainty of quantum measurement outcomes~\cite{Tapster94}. Unfortunately, imperfections in their components can make generated numbers predictable to some extent. This may be undetected by standard randomness tests~\cite{NIST}, and therefore exploited by an adversary~\cite{Colbeck07}.

A major breakthrough was the discovery of device-independent (DI) QRNGs, which permit to certify private randomness without making assumptions about the internal functioning of the devices~\cite{Colbeck07}. The problem is that current DI-QRNGs require very high detection efficiencies and produce private random numbers at low rate~\cite{PAMBMMOHLMM10}.

Another approach is the semi-device independent (SDI) QRNGs~\cite{PB11,rev1,Armin15}, in which no assumptions about the internal functioning of the QRNG are made, except that the dimension of the quantum system used is below a certain upper-bound. A typical SDI-QRNG consists of a device with two parts, $P$ and $M$, where $P$ prepares some quantum states that are then transmitted to $M$, where some quantum operations are performed producing outcomes which, after post-processing, are used as random bits. Previous works~\cite{Brunner15,MeasDI15,SourceDI16,SourceDI17,unamb17} consider SDI-QRNG protocols with the extra assumption that $P$ and $M$ are not correlated, which implies that $P$ and $M$ do not use shared randomness during the whole process of generation of random numbers. It has been shown that the presence of correlations between $P$ and $M$ makes DI-QRNG and SDI-QRNG protocols vulnerable to detection inefficiency attacks~\cite{attacks1,attacks2}.

There are three possible cases in which there is shared randomness between $P$ and $M$:
\begin{enumerate}[(i)]
	\item When correlations between $P$ and $M$ exist before the device is used. For example, if a common seed is stored by the adversary when the devices are built. Another example is when the same environment is shared by $P$ and $M$ and fluctuations in electrical power or local temperature affects them equally. This problem can avoided by employing several $P$s and $M$s paired randomly. Since the adversary cannot know in advance how they will be paired, then the adversary must resort to using common seed to all the parts, e.g., a synchronized timer. In this case, correlations between inputs and outputs will be observed, as discussed in~\cite{2s2r}.
	\item When a signal sent from an external synchronizer causes correlations between $P$ and $M$. This case can be avoided by invoking a standard assumption in all cryptographic schemes, namely, that $P$ and $M$ are inside a shielded laboratory, and thus a shared seed cannot be sent from the outside.
	\item When $P$ and $M$ correlate themselves during the execution of the protocol using communication. This can occur, e.g., in the following way. If $M$ is able to detect all photons sent from $P$, then, if $P$ decides not to send a photon for a certain round, this would be an indication for both $P$ and $M$ to reset their counters. In a more realistic case, when $M$ works with imperfect detectors, this synchronization attack strategy is still possible using a longer sequence of rounds without photons to reset their counters.
\end{enumerate}

In this work, we propose a SDI-QRNG protocol which offers a protection against these last attacks, and can be implemented with a very low detection efficiency. Its novelty is the introduction of a ``blocker'' that randomly stops the communication between the QRNG preparation and measurement stages with the purpose of destroying the correlations that may exist between them. Its relevance lies in the following features: (I)~It works with inefficient detectors, as required for real-world applications. (II)~No detailed knowledge of the internal functioning of the QRNG is needed. (III)~The protocol works even if the adversary has introduced shared randomness between $P$ and $M$.
In addition, to demonstrate the practicability of the protocol, we present an experimental realization of the protocol using weak coherent states and standard avalanche photo-detectors (APDs) providing an overall detection efficiency of 6\%.


\section{Protocol description}
\label{sec:2}

\begin{figure}[t]
	\centering
	\includegraphics[width=0.7\textwidth]{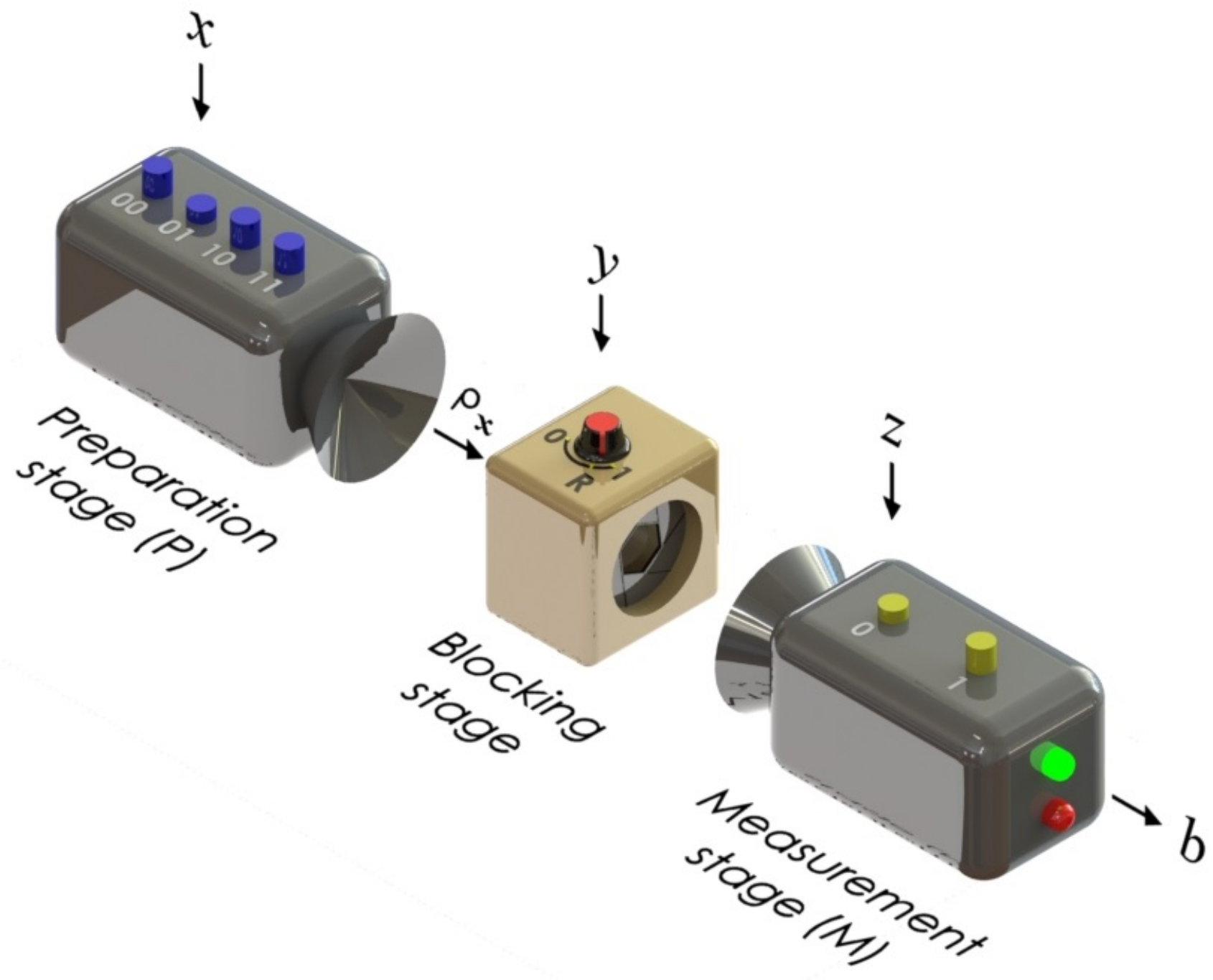}
	\caption{(Color online) Prepare-and-measure scenario of SDI protocols with a blocker for private randomness generation. It illustrates case with $X = \{00, 01, 10, 11\}$, $Z = \{0,1\}$, and $B = \{0,1\}$, the values of $b=0,1$ are represented by means of a top (green) or bottom (red) light, respectively, at the final stage.}
	\label{Fig1}
\end{figure}

Here, we overview the idea of the proposed protocol. Fig.~\ref{Fig1} shows the general scheme of SDI protocols and an extra blocker. It consists of three stages: the state preparation stage ($P$), the blocking stage, and the measurement stage ($M$).


In $P$, a quantum system $\rho_x$ is prepared depending on the input data $x \in X$. In the blocking stage, a blocker halts the transmitted system or allows it to go to the stage $M$, depending on another random input $y \in [0,1]$. The system is stopped if and only if $y \le R$, where $R \in [0,1]$ is a parameter of the blocker. In $M$ a random input $z \in Z$ is used to select a measurement which is then performed on the transmitted system. The outcome of the selected measurement is $b \in B \cup \{\emptyset\}$, where $b=\emptyset$ corresponds to the non-detecting events.

The SDI scheme considers $P$ and $M$ to be black boxes possibly built by an adversary. Their internal functioning is unknown to the user and they may even contain a malevolent agent. Still, the SDI approach assumes the following properties of the device:
\begin{enumerate}
	\item The dimension of the transmitted system is known.
	\item $P$ and $M$ do not receive any external signal from the adversary (i.e., the laboratory is shielded).
	\item $x,y,z$ are numbers independently produced that pass standard tests of randomness~\cite{NIST}, but are not private. They may be produced by imperfect QRNGs or taken from a public source of random bits.
\end{enumerate}

Each round of the protocol represents an event denoted by $(b|x,y,z)$. We take only the rounds with $y>R$ for our random string. We denote
\begin{equation}
	\mathbb{P}(b|x,z) \equiv \dfrac{\mathbb{P}(b|x,y > R,z)}{1 - \mathbb{P}(\emptyset|x,y > R,z)}, b \in B.
\end{equation}

To certify that the generated sequence of outputs is private and random, the user estimates the overall detection efficiency $\eta$ defined as
\begin{equation}
	\label{eq:eta}
	\eta \equiv 1 - \sum_{x \in X} \sum_{y \in Y} P_{XZ}(x,z) \mathbb{P}(\emptyset|x,y > R,z),
\end{equation}
and the value of the so-called certificate:
\begin{equation}
	\label{eq:certificate}
	W[\mathbb{P}] \equiv \sum_{x \in X} \sum_{y \in Y} \sum_{b \in B} \beta_{b,x,z} \mathbb{P}(b|x,z),
\end{equation}
with $P_{XZ}(x,z)$ being the probability distribution of settings in the considered case and the numbers $\beta_{b,x,z} \in \mathbb{R}$, defining a particular protocol. If $W[\mathbb{P}]$ is above a certain threshold that depends on $R$ and $\eta$, then the random sequence generated is considered as private. Otherwise the user aborts.

In order to prove the security of the proposed solution we consider the following characterization of the blocking and synchronization mechanism:
\begin{enumerate}[I.]
	\item The detection efficiency does not depend on the measurement setting used by $M$.
	\item Synchronization takes one round of communication and in this round no other information (e.g., about $x$) is transferred.
	\item $P$ sends synchronization signals with the same probability $\alpha \in [0, 1]$ in each round, i.e., the synchronization algorithm is memoryless.
	\item After blocking, $P$ and $M$ become uncorrelated until the next synchronization.
\end{enumerate}
We thus consider three separate cases of runs: runs not synchronized (i.e. ones in that $P$ and $M$ are not correlated), runs used for synchronization, and runs synchronized (and not containing the synchronization signal). In practice, it is probable that any form of synchronization will require more than e.g. a single qubit, so the second assumption favors the adversary.

These assumptions significantly simplify the calculation of the amount of certified randomness. In a more general approach, one may consider sophisticated synchronization strategies, in which synchronization signals are being sent according to some patterns or take into account correlations which remains to some extent even after the blocking\footnote{For example, the internal counters of $P$ and $M$ differ by $n$ rounds with some probability, depending on $n$ and the blocking rate.}.

Let $c_l$ be the maximal value of the certificate~\eqref{eq:certificate} if the transmitted system is classical, and $c_q$ if it is quantum, for some fixed dimension. We assume $c_l < c_q$. As mentioned above, if the detection efficiency~\eqref{eq:eta} is less than $1$ and the parts are synchronized, then there are strategies able to mimic higher values of the certificate~\eqref{eq:certificate} than allowed by quantum mechanics. Let $c_s(\eta)$ be the maximal value of the certificate~\eqref{eq:certificate} for a given detection efficiency $\eta$. Similarly, let $c_r$ be the maximal value of~\eqref{eq:certificate} when no information is transmitted and $M$ calculates the outcome $b$ depending on the input $z$. This is the case when the transmitted system contains a synchronization signal, cf.~the assumption~II. We have:
\begin{equation}
	\bigforall_{0 < \eta \leq 1} 0 \leq c_r \leq c_l < c_q \leq c_s(\eta) \leq 1.
\end{equation}


\section{Main result: Blocking protocol theorem}
\label{sec:3}


A common measure of randomness generated by QRNGs is the min-entropy. One defines the so called guessing probability of a discrete distribution $\mathbb{P}$ as $P_{guess}[\mathbb{P}] \equiv \max_{x \in \text{supp}(\mathbb{P})} \mathbb{P}(x)$. The min-entropy is then defined as $H_{\infty}[\mathbb{P}] \equiv -\log_2 P_{guess}[\mathbb{P}]$~\cite{minEntr}.
For a conditional distribution $\mathbb{P}_{\Phi|\Psi}$, $\psi$ distributed with $P_{\Psi}$ one~\cite{Armin15} defines:
\be
	P_{guess}[\mathbb{P}_{\Phi|\Psi}] = \max_{\bm{g}: \Psi \to \Phi} \Biggl\{ \sum_{\psi \in \Psi} P_{\Psi}(\psi) \mathbb{P}(\bm{g}(\psi)|\psi) \Biggr\}.
\ee

More generally, if $\mathbb{P}$ is an (unnormalized) mixture of distributions $\mathbb{P}_i$ (conditional or not) with frequencies $\{\omega_i\}$, $\sum_i \omega_i \leq 1$, we define $H_{\infty}[\mathbb{P}] \equiv - \sum_i \omega_i H_{\infty}[\mathbb{P}_i]$.

The min-entropy of a set $\mathcal{Q} = \{\mathbb{P}\}$ certified by a value $p$ of a certificate $W: \mathcal{Q} \to \mathbb{R}$ is
\be
	H_{\infty}\left[ \mathcal{Q}, W, p \right] \equiv \inf_{\substack{\mathbb{P} \in \mathcal{Q}}} \{ H_{\infty}[\mathbb{P}] : W[\mathbb{P}] = p \}.
\ee
If $W[\mathbb{P}] > c_l$, then it is not possible to reproduce the probability distribution with local hidden variables models, thus $H_{\infty}[\mathbb{P}] > 0$.

We denote the certified randomness in an experiment with the detection efficiency $\eta$, the blocking rate $R$ and the observed certificate value $p$ by $H_{\infty}(R,\eta,p)$. We provide details of this function in Appendix~A. The set of all conditional distributions possible to be obtained in it for a given $\eta$ differ between runs without and with correlations, and runs with the synchronization signal, denoted $\mathcal{Q}_\text{asyn}(\eta)$, $\mathcal{Q}_\text{syn}(\eta)$ and $\mathcal{Q}_\text{sig}(\eta)$, with frequencies $\omega_\text{asyn}$, $\omega_\text{syn}$ and $\omega_\text{sig}$, respectively. $H_{\infty}(R,\eta,p)$ is thus the min-entropy of a mixture of distributions from these sets with joint constraints on the average value of $p$ and $\eta$.

The main result of this work is the following:
\begin{trm}
	\label{trm:main}
	If for an SDI-QRNG protocol using a certificate~\eqref{eq:certificate} we obtain in an experiment the value $p > c_l$, then for any detection efficiency $\eta > 0$ there exists a blocking rate $R$ providing protection against detection loophole attacks, i.e.,
	\be
		\bigforall_{p > c_l} \bigforall{\eta > 0} \bigexists_{R \in [0,1)} H_{\infty}(R,\eta,p) > 0.
	\ee
\end{trm}
We present the proof in Appendix~B.


\section{Experiment}
\label{sec:4}

\begin{figure}[t]
	\centering
	\includegraphics[width=0.75\textwidth]{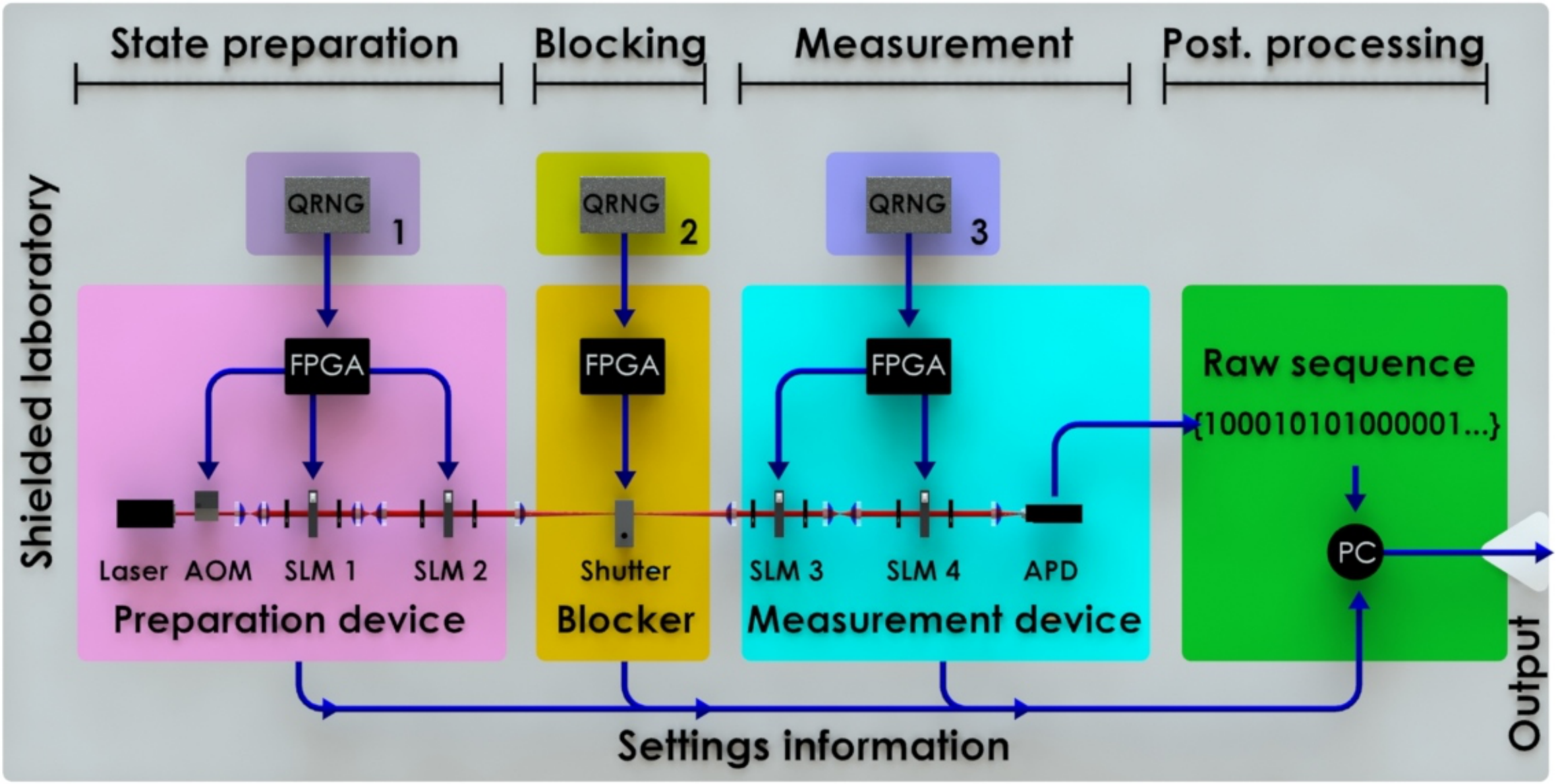}
	\caption{(Color online) Experimental setup (see text for details).}
	\label{Fig45}
\end{figure}

Our experimental implementation of the protocol with a blocker is shown schematically in Fig.~\ref{Fig45}. As the sources of $x$, $y$, and $z$ we use three commercial QRNGs (IDQ Quantis). They passed standard tests of randomness, but no assumptions about their privacy is made.

A field programmable gate array (FPGA) in $P$ produces an electrical synchronization signal, which also drives an acousto-optical modulator (AOM) producing attenuated optical pulses (weak coherent states) from a continuous laser operating with a center wavelength of 690 nm. These optical pulses are then sent through a sequence of four spatial light modulators (SLMs)~\cite{Grier03}. Sets of lenses are employed to project the image of one SLM onto the next one.

The assumption that the quantum system prepared in $P$ and measured in $M$ is two dimensional is addressed in our experiment in the following way. The employed average photon number per optical pulse was set to $\mu = 0.66$ such that approximately 71\% of the non-null pulses contain only one photon. To define the two dimensional quantum systems we use the linear transverse momentum degree of freedom of the photons transmitted by the SLMs~\cite{LVNRS09}. This is done by projecting masks with only two paths available for the photon transmission in the liquid crystal displays of the SLMs~\cite{Neves05,ref1,ref2,ref3,ref4,ref5,refDoubleslit}.

The qubit state preparation in $P$ (and the projections in $M$) is implemented using SLM~1 and SLM~2 [SLM~3, SLM~4 (and an APD)] working with amplitude-only and phase-only modulation, respectively~\cite{LNGGNDVS11,ECGNSXL13,ref6}. The real and imaginary parts of the generated and measured states are set by adjusting the grey level of the pixels on the SLMs. In our demonstration we set these states to maximise the value of $W[\mathbb{P}]$ (described in Appendix~C).

The repetition rate of the attenuated optical pulses is set to 30~Hz, which is the limit of the employed SLMs. The applied modulation in each SLM is triggered by the sync signal. An internal delay in respect to the AOM in the FPGAs is used to ensure that the SLMs in $P$ and $M$ are properly set by the time each optical pulse is sent. In each round, pre-determined modulations are applied to the SLMs by their corresponding FPGAs based on the numbers produced by each QRNG. The optical blocker placed between them is a commercial shutter and is controlled by a third FPGA unit, fed by another QRNG. The blocker's electronics also receive the sync signal from $P$. For each round of the experiment, the blocker's FPGA unit randomly blocks the optical pulse, with an adjustable probability. The overall detection efficiency $\eta$ (including losses at the measurement stage and detection probability of the single-photon detector) is 6\%.

Note that the stages $P$, $B$, and $M$ rely on the sync signal, which represents the action of choosing a random input $x$ in $P$, $y$ in $B$ and $z$ in $M$. If synchronization between $P$ and $M$ is achieved, then an adversary's agent in $M$ could use the sync signal to count rounds and prevent the loss of synchronization. However, this attack can easily be counteracted by the user: If he randomly sends fake signals to $M$ between the sync signals then synchronization will be again required for the adversary, which can be detected by the protocol.


In our implementation a blocking rate $R = 0.99$ was employed. We obtained $W[\mathbb{P}] = 0.824 \pm 0.015$ with data shown in Tab.~1 
in Appendix~C. A direct calculation using a relaxed formula for $H_{\infty}(R,\eta,p)$ for the considered setup (see~(C19) shows that the key generation rate is at least $0.009448$ bits of min-entropy per photon passing the blocker mechanism (or $0.000094$ bits per emitted photon).


\section*{Conclusions}
\label{sec:5}


We have introduced and experimentally implemented a SDI-QRNG protocol protected against attacks based on the detection efficiency loophole. In the experimental implementation, we have reported a private random bit rate generation of $0.000094$ per emitted photon with a detection efficiency of 6\%.

Unlike previous protocols which make the assumption that the measuring \cite{MeasDI15} or the preparing parts \cite{SourceDI16,SourceDI17} are trustworthy or that there are not correlations between them \cite{unamb17}, our protocol does not need to make any of these assumptions.

Unlike a recent protocol~\cite{highSpeed18} that does not make these assumptions but requires detection efficiencies above $78\%$ and produces relatively low randomness, our protocol works with much lower detection efficiencies and produces more randomness. For example, while the protocol in \cite{highSpeed18}
produces $0.00114$ bits per round, our protocol certifies $0.071553$ bits per emitted photon with $\eta=0.78$ and $R=0.3$.

We believe that our results pave the way towards a new generation of practical and secure SDI-QRNGs.


\section*{Acknowledgments}


This work was supported by Fondecyt~1160400, Fondecyt~11150324, Fondecyt~11150325, Fondecyt~1150101, and Millennium Institute for Research in Optics, MIRO. J.F.B. acknowledges support from Fondecyt~3170307. J.C. acknowledges support from Fondecyt~3170596. A.C. acknowledges support from the Ministry of Science, Innovation and Universities (MICIU) Grant No.~FIS2017-89609-P with FEDER funds, the Conserjer\'{\i}a de Conocimiento, Investigaci\'on y Universidad, Junta de Andaluc\'{\i}a and European Regional Development Fund (ERDF) Grant No.~SOMM17/6105/UGR, and the Knut and Alice Wallenberg Foundation project ``Photonic Quantum Information.'' G.B.X. acknowledges Ceniit Link\"oping University and the Swedish Research Council (VR 2017-04470) for financial support. P.M. and M.P are supported by a National Science Centre (NCN) grant 2014/14/E/ST2/00020 and FNP programme First TEAM (Grant No. First TEAM/2016-1/5), and P.M. by a DS Programs of the Faculty of Electronics, Telecommunications and Informatics, Gda\'nsk University of Technology. P.M. thanks Krystyna Witalewska for help during the time of writing the manuscript. The optimizations have been performed using OCTAVE~\cite{octave} with SeDuMi solver~\cite{SeDuMi} and YALMIP toolbox~\cite{yalmip}.




\section{Appendix A: Guessing probability, min-entropy, and randomness certification}


The distribution of lengths of series of systems sent and not blocked by the blocking mechanism at a rate $R$ is given by
\be
	P(R, k) = (1 - R)^{k-1} R,
\ee
where $k \geq 1$ is the number systems in a series.

Let us denote by $\alpha$ the ratio between the number of systems containing synchronization signal sent by $P$ and the total number of systems sent by $P$, cf.~the assumption~III.

In a given series of transmitted systems of length $k$, the average number of photons before the first synchronizing system within that series is
\be
	\label{eq:AvgBeforeSynch}
	\ba
		F_\text{asyn}(\alpha, k) &= \left[ \sum_{i=0}^{k} i (1 - \alpha)^i \alpha \right] + k (1 - \alpha)^k \\
		&= \frac{1 - \alpha}{\alpha} \left(1 - (1 - \alpha)^k\right).
	\ea
\ee
Using~\eqref{eq:AvgBeforeSynch}, we get that the probability that a particular photon received by $M$ is not synchronized is equal
\be
	\label{eq:ProbBeforeSynch}
	\ba
		P_\text{asyn}(R, \alpha) &= \sum_{k=1}^{\infty} P(R, k) \frac{F_\text{asyn}(\alpha, k)}{k} \\
		&= \frac{1 - \alpha}{\alpha} \frac{R}{1 - R} \ln \left( \frac{\alpha + R - \alpha R}{R} \right),
	\ea
\ee
for $\alpha > 0$ and $P_\text{asyn}(R, 0) = 1$. Directly from the definition of $\alpha$, the probability that the received system is a synchronization signal is
\be
	\label{eq:ProbSynching}
	P_\text{sig}(\alpha) = \alpha.
\ee
The probability that a received system is synchronized and not carrying a synchronization signal is
\be
	\label{eq:ProbSynchNotSynching}
	P_\text{syn}(R, \alpha) = 1 - P_\text{asyn}(R, \alpha) - P_\text{sig}(\alpha),
\ee
for $\alpha > 0$ and $P_\text{syn}(R, 0) = 0$. One can easily calculate that
	$\bigforall_{\alpha \in [0,1]} \lim_{R \to 1^{-}} P_\text{syn}(R, \alpha) = 0$.
Let us define $F_\text{syn}(R) \equiv \max_{\alpha \in [0,1]} P_\text{syn}(R, \alpha)$. Then,
\be
	\label{eq:Fsyn}
	\lim_{R \to 1^{-}} F_\text{syn}(R) = 0.
\ee

Let us denote:
\begin{subequations}
	\be
		\label{eq:Hasyn}
		H_{\infty}^\text{asyn}(p,\eta) \equiv H_{\infty}\left[ \mathcal{Q}_\text{asyn}(\eta), W, p \right],
	\ee
	\be
		\label{eq:Hsyn}
		H_{\infty}^\text{syn}(p,\eta) \equiv H_{\infty}\left[ \mathcal{Q}_\text{syn}(\eta), W, p \right].
	\ee
\end{subequations}
Runs without synchronization on average produce min-entropy given by~\eqref{eq:Hasyn}. Let $p_\text{asyn}$ be the average value of the certificate~(3), and $\eta_\text{asyn}$ be the detection efficiency~(2) restricted to these rounds. Similarly let $p_\text{sig}$ and $\eta_\text{sig}$ refer to the runs containing synchronization signal, and $p_\text{syn}$ and $\eta_\text{syn}$ for the synchronized runs without the signal. These values cannot be observed individually by user, but each of them has impact on the observed values of~(3) and~(2). Let $\vec{p} \equiv (p_\text{asyn},p_\text{sig},p_\text{syn})$ and $\vec{\eta} \equiv (\eta_\text{asyn},\eta_\text{sig},\eta_\text{syn})$. Note that $p_\text{asyn} \leq c_q$, $p_\text{sig} \leq c_r$ and $p_\text{syn} \leq c_s(\eta_\text{syn})$. Let
\be
	\ba
		& \omega_\text{asyn} \equiv \eta_\text{asyn} P_\text{asyn}(R,\alpha), \\
		& \omega_\text{sig} \equiv \eta_\text{sig} P_\text{sig}(\alpha), \\
		& \omega_\text{syn} \equiv \eta_\text{syn} P_\text{syn}(R,\alpha), \text{ and:}
	\ea
\ee
\begin{subequations}
	\be
		\ba
			H_{\infty}^\text{fun} &\left( R,\alpha,\vec{p},\vec{\eta} \right) \equiv \\ & \omega_\text{asyn} H_{\infty}^\text{asyn}(p_\text{asyn}, \eta_\text{asyn}) + \omega_\text{syn} H_{\infty}^\text{syn}(p_\text{syn},\eta_\text{syn}),
		\ea
	\ee
	\be
		\label{eq:pFun}
		p^\text{fun} \left( R,\alpha,\vec{p},\vec{\eta} \right) \equiv p_\text{asyn} \omega_\text{asyn} + p_\text{sig} \omega_\text{sig} + p_\text{syn} \omega_\text{syn},
	\ee
	\be
		\label{eq:etaFun}
		\eta^\text{fun} \left( R,\alpha,\vec{\eta}\right) \equiv \omega_\text{asyn} + \omega_\text{sig} + \omega_\text{syn}.
	\ee
\end{subequations}

Using the above expression we formulate a lower bound on the certified min-entropy for the blocking parameter $R$, the observed value $p$ of the certificate~(3) and the observed value $\eta$ of the detection efficiency~(2):
\be
	\label{eq:Hinf}
	\ba
		H_{\infty}(R,\eta,p) \equiv & \underset{\substack{\alpha \in [0,1] \\ \eta_\text{asyn},\eta_\text{sig},\eta_\text{syn} \in [0,1] \\ p_\text{asyn} \in [0,c_q] \\ p_\text{sig} \in [0,c_r] \\ p_\text{syn} \in [0,c_s(\eta_\text{syn})]}}{\text{minimize}} & H_{\infty}^\text{fun} \left( R,\alpha,\vec{p},\vec{\eta} \right) \\
		\null & \quad \text{subject to } \quad & p^\text{fun} \left( R,\alpha,\vec{p},\vec{\eta} \right) = \eta p, \\
		\null & \null & \eta^\text{fun} \left( R,\alpha,\vec{\eta}\right) = \eta.
	\ea
\ee

\section{Appendix B: Proof of the blocking protocol theorem}


From~\eqref{eq:Hinf} it follows that to have $H_{\infty}(R,\eta,p) = 0$ we need
\begin{subequations}
	\be
		\label{eq:noHinfAsyn}
		H_{\infty}^\text{asyn}(p_\text{asyn},\eta_\text{asyn}) = 0, \text{ or}
	\ee
	\be
		\label{eq:noPAsyn}
		\omega_\text{asyn} = 0.
	\ee
\end{subequations}
For~\eqref{eq:noHinfAsyn} to hold we need $p_\text{asyn} \leq c_l$. Knowing this we relax the constraints in the optimization problem~\eqref{eq:Hinf}:
\begin{subequations}
	\label{eqs:relax1}
	\be
		\label{eq:relax1etaP}
		c_l \omega_\text{asyn} + c_r \omega_\text{sig} + P_\text{syn}(R,\alpha) \geq \eta p,
	\ee
	\be
		\label{eq:relax1eta}
		\omega_\text{asyn} + \omega_\text{sig} \leq \eta.
	\ee
\end{subequations}
Using~\eqref{eqs:relax1} and $c_l \geq c_r$, we get
\be
	P_\text{syn}(R,\alpha) \geq \eta (p - c_l) \equiv \epsilon > 0.
\ee
From~\eqref{eq:Fsyn} we know there exists $R < 1$ such that $P_\text{syn}(R,\alpha) < \epsilon$. The contradiction shows that~\eqref{eq:noHinfAsyn} cannot be satisfied for all $R$.

What remains for the proof is to show that also~\eqref{eq:noPAsyn} is not true. To show this consider the following relaxation of constraints in~\eqref{eq:Hinf}:	
\be
	\label{eqs:relax3}
	c_q \omega_\text{asyn} + c_r \omega_\text{sig} + F_\text{syn}(R) \geq \eta p,
\ee
and~\eqref{eq:relax1eta}. Substituting the latter into~\eqref{eqs:relax3} we get
\be
	\label{eq:PasynLowerBound}
	\omega_\text{asyn} \geq \frac{\eta \cdot (p - c_r) - F_\text{syn}(R)}{c_q - c_r}.
\ee
Since, for some blocking rate $R < 1$, the value of $F_\text{syn}(R)$ is arbitrary small and $p - c_r > 0$, we see that also~\eqref{eq:noPAsyn} is not satisfied. Therefore, there exists a blocking rate $R$ such that $H_{\infty}(R,\eta,p) > 0$.



\section{Appendix C: Randomness of $2 \to 1$ quantum random access code}


A common example~\cite{PB11,LYWZWCGH11,MLP14} of a certificate~(3) is based on the so called $2 \to 1$ quantum random access code~\cite{ANTV02,QRACs} in dimension $2$:
\be
	\label{eq:certificateQRAC21}
	W^{2 \to 1}\left[\mathbb{P}\right] \equiv \frac{1}{8} \sum_{x \in \{00,01,10,11\} } \sum_{z \in \{1,2\} } \mathbb{P}(b=x_z|x,z).
\ee
Results of the experimental implementation of this QRAC are shown in Tab.~\ref{tab:qracData}.
\begin{table}[t]
\caption{Observed experimental probabilities $\mathbb{P}(b|x,z)$ with error bars. The bold values refers to QRAC successes (their maximal possible average value is $0.5 + \frac{\sqrt{2}}{4} \approx 0.85355$). Please note that experimental losses, together with the average photon number per pulse µ, do affect the observed success probabilities reported here. However, as we have demonstrated recently \cite{ref3}, in a QRAC the decrease in the average success probability is only linear in the term µν (where ν represents the losses), Thus, the effect of multiphoton events while using µ=0.66, and an overall detection efficiency of $6\%$ is minimal. This can be corroborated by noting that the obtained results are close to the ideal one.}
\label{tab:qracData}
	\begin{tabular}{|r|c|c|}
		\hline $z = 0$ & $b = 0$ & $b = 1$ \\ \hline
		$x = 00$ & $\mathbf{0.850016 \pm 0.026011}$ & $0.149984 \pm 0.007815$ \\ \hline
		$x = 01$ & $\mathbf{0.858255 \pm 0.029392}$ & $0.141745 \pm 0.008488$ \\ \hline
		$x = 10$ & $0.145144 \pm 0.007938$ & $\mathbf{0.854856 \pm 0.027039}$ \\ \hline
		$x = 11$ & $0.146975 \pm 0.007632$ & $\mathbf{0.853025 \pm 0.025767}$ \\ \hline
		\hline $z = 1$ & $b = 0$ & $b = 1$ \\ \hline
		$x = 00$ & $\mathbf{0.813494 \pm 0.020024}$ & $0.186506 \pm 0.007073$ \\ \hline
		$x = 01$ & $0.148772 \pm 0.006703$ & $\mathbf{0.851228 \pm 0.022439}$ \\ \hline
		$x = 10$ & $\mathbf{0.820518 \pm 0.020665}$ & $0.179482 \pm 0.007086$ \\ \hline
		$x = 11$ & $0.161361 \pm 0.006431$ & $\mathbf{0.838639 \pm 0.020305}$ \\ \hline
	\end{tabular}
\end{table}

To calculate the amount of min-entropy generated by the proposed protocol, we performed optimization over the set of all probability distributions allowed by quantum mechanics in a scenario in which $P$ prepares one of $4$ quantum states of dimension $2$ and sends it to $M$ performing one of two possible binary measurements. We used the see-saw technique~\cite{seesaw1,seesaw2} of semi-definite programming~\cite{SDP} and computed the following quantity:
\be
	\label{eq:optSeesaw}
	\ba
		h(p) & \equiv & \underset{ \{ \rho_i \}, \{ M \}, \bm{g} }{\text{minimize}} \quad & -\log_2 \left(\frac{1}{8} \sum_{x,z} P(\bm{g}(x,z) | x, z) \right) \\
		\null & \null & \text{subject to } \quad & \frac{1}{8} \sum_{x,z} P(x_z | x, z) = p, \\
		\null & \null & \null & P(b|x,z) = \Tr \left( \rho_x M^z_b \right),
	\ea
\ee
where $x \in \{0,1\}^2$, $z \in \{0,1\}$, $\{ \rho_i \} = \{ \rho_{00},\rho_{01},\rho_{10},\rho_{11} \}$ are states, $\{ M \} = \{ \{M^0_0,M^0_1\}, \{M^1_0,M^1_1\} \}$ are measurements on a Hilbert space of dimension $2$, and $\bm{g}: \{0,1\}^3 \rightarrow \{0,1\}$ is a possible guessing strategy when $x$ and $z$ are known. We took $P_{XZ}(x,z) = \frac{1}{\Ab{X} \Ab{Z}} = \frac{1}{8}$. In this case, $c_l = \frac{3}{4}$ and $c_q = \frac{1}{2}+\frac{\sqrt{2}}{4}$. Since without correlations the detection loophole cannot be exploited, we have simply
\be
	H_{\infty}^\text{asyn}(p,\eta) = h(p).
\ee

When $\eta = 1$ and $P$ and $M$ are synchronized, then they can use shared randomness to mix their strategies. The randomness in this case is given by a convex combination of the randomness obtained in each of the mixed strategies, i.e. by $\Hull [h] (p)$, where the functional $\Hull [\cdot]$ transforms a function to its convex hull (a maximal lower-bounding convex function). The following bound for $2 \to 1$~QRAC holds:
\be
	\label{eq:HSR2to1}
	H^{2 \to 1}(p) \equiv k \cdot \left(p - \frac{3}{4}\right) \leq \Hull [h] (p),
\ee
with $k \equiv 4 \cdot \left(\sqrt{2} + 1\right) \log_2\left(4 - 2 \sqrt{2}\right)$. The bound can be obtained directly by solving the optimization problem~\eqref{eq:optSeesaw} and taking its convex hull. It is also trivial that
\be
	\label{eq:Hasyn2to1}
	H_{\infty}^\text{asyn}(p,\eta) = h(p) \geq H^{2 \to 1}(p),
\ee
as $H^{2 \to 1}$ gives the malevolent constructor a power of using shared randomness for mixing guessing strategies (but not for detection loophole attack).

Now, we give the explicit formula for~\eqref{eq:Hsyn} for the considered protocol. $M$ is allowed not to click in $1-\eta$ part of rounds, and using detection efficiency loophole it can mimic a higher value of the certificate~\eqref{eq:certificateQRAC21}. The method is the following.

If $\eta > \frac{1}{2}$ then in $2(1-\eta)$ part of rounds $P$ is encoding one bit from the input, $x_0$ or $x_1$ with equal ratio. The choice which bit to encode is guided by shared randomness. If the input $z$ matches the encoded bit, $M$ measures the qubit and satisfies the certificate~\eqref{eq:certificateQRAC21} with probability $1$. Otherwise $M$ outputs $\emptyset$. In the remaining $2\eta-1$ part of rounds, $P$ and $M$ use the states and measurements referring to the value of the certificate~\eqref{eq:certificateQRAC21} equal to some $q \in [c_l, c_q]$. The observed average value of the certificate in such a strategy is
\be
	\label{eq:PSynFun}
	\frac{1}{\eta} \left[ (2\eta-1) q + (1-\eta) \right].
\ee
The detection loophole allows to achieve up to $\frac{2\eta + \sqrt{2} - 1}{2 \sqrt{2} \eta} \leq 1$ with $q = c_q$. The value of \eqref{eq:PSynFun} equals $p \leq c_s({\eta})$ if and only if $q = \frac{\eta (1+p) - 1}{2\eta -1}$.

If $\eta \leq \frac{1}{2}$ then the malevolent vendor can use the above strategy for all inputs and attain $W^{2 \to 1}\left[\mathbb{P}\right] = 1$. Thus
\be
	c_s(\eta) = \min \left( \frac{2\eta + \sqrt{2} - 1}{2 \sqrt{2} \eta} , 1 \right).
\ee

The average min-entropy generated in rounds with detections is thus given by
\be
	\label{eq:Hsyn2to1}
	H_{\infty}^\text{syn}(p,\eta) =
	\begin{cases}
		\frac{2\eta - 1}{\eta} \cdot H_{\infty}^\text{SR}\left(\frac{\eta (1+p) - 1}{2\eta -1}\right) & \text{if } \eta > \frac{1}{2}, \\
		0 & \text{if } \eta \leq \frac{1}{2}.
	\end{cases}
\ee


Let us now give an explicit formula for a lower bound on min-entropy in this scenario. First, let us define
\be
	\ba
		h^{2 \to 1}&\left(R, \alpha, p_\text{asyn}, q, \eta_\text{asyn}, \eta_\text{syn}\right) \equiv \\
		&\begin{cases}
			\omega_\text{asyn} H^{2 \to 1}(p_\text{asyn}) + & \text{for } \eta_\text{syn} > \frac{1}{2}, \\
			\quad (2 \eta_\text{asyn} - 1) P_\text{syn}(R, \alpha) H^{2 \to 1}(q) & \\
			\omega_\text{asyn} H^{2 \to 1}(p_\text{asyn}) & \text{for } \eta_\text{syn} \leq \frac{1}{2},
		\end{cases}
	\ea
\ee
and
\be
	\label{eq:pOfq}
	p_\text{syn}^\text{fun}(q, \eta_\text{syn}) \equiv
	\begin{cases}
		2 q - 1 + \frac{1-q}{\eta_\text{syn}} & \text{for } \eta_\text{syn} > \frac{1}{2}, \\
		1 & \text{for } \eta_\text{syn} \leq \frac{1}{2}.
	\end{cases}
\ee
From~\eqref{eq:HSR2to1}, \eqref{eq:Hasyn2to1}, and~\eqref{eq:Hsyn2to1} we get a bound
\be
	\label{eq:HfunH2to1bound}
	H_{\infty}^\text{fun} \left( R,\alpha,\vec{p},\vec{\eta} \right) \geq h^{2 \to 1}\left(R, \alpha, p_\text{asyn}. \eta_\text{asyn}, \eta_\text{syn}\right).
\ee
Substituting~\eqref{eq:HfunH2to1bound} by~\eqref{eq:Hinf} and using~\eqref{eq:pOfq}, we obtain
\be
	\label{eq:HinfBound}
	\ba
		H_{\infty}(R,\eta,p) \geq & \underset{\substack{\alpha \in [0,1] \\ \eta_\text{asyn},\eta_\text{sig},\eta_\text{syn} \in [0,1] \\ p_\text{asyn},q \in [\frac{3}{4},c_q] \\ }}{\text{minimize}} h^{2 \to 1} \left(R, \alpha, p_\text{asyn}, q, \eta_\text{asyn}, \eta_\text{syn}\right) \\
		\null & \quad \text{subject to } \quad q^\text{fun} \left( R,\alpha,p_\text{asyn},q,\vec{\eta} \right) \geq \eta p, \\
		\null & \quad \quad \quad \quad \quad \quad \quad \eta^\text{fun} \left( R,\alpha,\vec{\eta}\right) = \eta,
	\ea
\ee
where $q^\text{fun}$, cf.~\eqref{eq:pFun}, is given by
\be
	\label{eq:pFunq}
	\ba
		q^\text{fun} \left( R,\alpha,p_\text{asyn},q,\vec{\eta} \right) \equiv & p_\text{asyn} \omega_\text{asyn} + \frac{1}{2} \omega_\text{sig} \\
		&+ p_\text{syn}^\text{fun}(q, \eta_\text{syn}) \omega_\text{syn}.
	\ea
\ee

Let us denote $x \equiv \omega_\text{asyn}$ and $y \equiv \omega_\text{syn}$.
Now, use the constraint $\eta^\text{fun} \left( R,\alpha,\vec{\eta}\right) = \eta$ to eliminate $\eta_\text{sig}$ and employ the bound~\eqref{eq:HSR2to1}. We separate further analysis in two cases.

In the case with $\eta_\text{syn} \in [0, \frac{1}{2}]$ we have $y \leq \frac{1}{2} P_\text{syn}(R,\alpha)$. Let us denote for this case $z^{-} \equiv x \cdot \left(p_\text{asyn} - \frac{3}{4}\right)$. Then~\eqref{eq:HinfBound} transforms to
\be
	\label{eq:linearOptim1}
	H_{\infty}^{-}(R,\eta,p) \geq \underset{\substack{\alpha \in [0,1] \\ (x,y,z^{-}) \in A^{-}(R,\eta,p,\alpha)}}{\text{minimize}} k z^{-},
\ee
where $A^{-}(R,\eta,p,\alpha) \subseteq \mathcal{R}^3$ is defined by the following linear constraints:
\be
	\ba
	& 0 \leq x \leq P_\text{asyn}(R,\alpha), \\
	& 0 \leq y \leq \frac{1}{2} P_\text{syn}(R,\alpha), \\
	& 0 \leq z^{-} \leq \left(c_q - \frac{3}{4}\right) \cdot x, \\
	& \eta - P_\text{sig}(\alpha) \leq x + y \leq \eta, \\
	& \eta \cdot \left(p - \frac{1}{2}\right) - \frac{1}{4} x - \frac{1}{2} y \leq z^{-}.
	\ea
\ee

In the case with $\eta > \frac{1}{2}$ we have $y \geq \frac{1}{2} P_\text{syn}(R,\alpha)$. Let us now denote
\be
	z^{+} \equiv x \cdot \left(p_\text{asyn} - \frac{3}{4}\right) + (2 y - P_\text{asyn}(R,\alpha)) \cdot \left(q - \frac{3}{4}\right).
\ee
As in the previous case,~\eqref{eq:HinfBound} transforms to
\be
	\label{eq:linearOptim2}
	H_{\infty}^{+}(R,\eta,p) \geq \underset{\substack{\alpha \in [0,1] \\ (x,y,z^{+}) \in A^{+}(R,\eta,p,\alpha)}}{\text{minimize}} k z^{+},
\ee
where $A^{+}(R,\eta,p,\alpha) \subseteq \mathcal{R}^3$ is defined by the following linear constraints:
\be
	\ba
		& 0 \leq x \leq P_\text{asyn}(R,\alpha), \\
		& \frac{1}{2} P_\text{syn}(R,\alpha) \leq y \leq P_\text{syn}(R,\alpha), \\
		& 0 \leq z^{+} \leq \left(c_q - \frac{3}{4}\right) \cdot \left[x + 2 y - P_\text{syn}(R,\alpha)\right], \\
		& \eta - P_\text{sig}(\alpha) \leq x + y \leq \eta, \\
		& \eta \cdot \left(p - \frac{1}{2}\right) - \frac{1}{4} x - \frac{1}{4} P_\text{syn}(R,\alpha) \leq z^{+}.
	\ea
\ee

It is easy to see that for fixed $\alpha \in [0,1]$ the internal optimization in~\eqref{eq:linearOptim1} and~\eqref{eq:linearOptim2} is a linear program. Thus we derive the following formula for $H_{\infty}(R,\eta,p)$ for the this scenario:
\be
	\label{eq:HinfRelax}
	H_{\infty}(R,\eta,p) \geq \min \left( H_{\infty}^{-}(R,\eta,p), H_{\infty}^{+}(R,\eta,p) \right).
\ee


\end{document}